\RequirePackage{tikz}
\documentclass[journal,twoside,web]{ieeecolor}
\usepackage{orcidlink}
\usepackage{generic}
\usepackage[
  backend=biber,
  style=ieee,
  maxbibnames=1,
  minbibnames=1,
  doi=true,
  url=false,
  eprint=false,
  isbn=false
]{biblatex}
\usepackage{amsmath,amssymb,amsfonts}
\usepackage{array}
\usepackage{algorithmic}
\usepackage{graphicx}
\usepackage{subcaption}
\usepackage{algorithm,algorithmic}
\usepackage{hyperref}
\usepackage[capitalize]{cleveref}
\usepackage{tabularx}
\usepackage{pifont}
\usepackage{booktabs}
\usepackage{comment}
\usepackage{bookmark}
\usepackage{siunitx}

\providecommand{\refname}{References}
\makeatletter
\let\oldthebibliography\thebibliography
\renewcommand{\thebibliography}[1]{\oldthebibliography{10}}
\makeatother

\usepackage{textcomp}
\def\BibTeX{{\rm B\kern-.05em{\sc i\kern-.025em b}\kern-.08em
    T\kern-.1667em\lower.7ex\hbox{E}\kern-.125emX}}
\makeatletter
\renewcommand{\ps@titlepagestyle}{%
  \def\@oddfoot{}\def\@evenfoot{}%
  \def\@oddhead{\begin{tabular}{@{}p{43pc}@{}}\\[-24pt]\vbox{\hsize43pc\scriptsize\textsf\leftmark \hfill \textsf\thepage\hbox{}}\\[-19pt]%
        \vbox{\color{subsectioncolor}\hrule height1pt width43pc depth0pt} \end{tabular}}%
  \def\@evenhead{\begin{tabular}{@{}p{43pc}@{}}\\[-24pt]\vbox{\hsize43pc\scriptsize\textsf\thepage \hfill \textsf\leftmark\hbox{}}\\[-19pt]%
        \vbox{\color{subsectioncolor}\hrule height1pt width43pc depth0pt} \end{tabular}}%
}
\makeatother

\begin{document}
\title{Evaluation of Grid-based Uncertainty Propagation for Collaborative Self-Calibration in Indoor Positioning Systems}
\author{Paul Schwarzbach\orcidlink{0000-0002-1091-782X}$^{\dagger}$ and Andrea Jung\orcidlink{0000-0003-1019-6134}$^{\dagger}$
\thanks{Corresponding author: Paul Schwarzbach. ${\dagger}$ Authors contributed equally.}
\thanks{The authors are with the Institute of Traffic Telematics, TUD Dresden University of Technology, 01069 Dresden, Germany (e-mail: firstname.lastname@tu-dresden.de). }}

\maketitle
\begin{abstract}
Radio-based localization systems conventionally require stationary reference points (e.g. anchors) with precisely surveyed positions, making deployment time-consuming and costly. This paper presents an empirical evaluation of collaborative self-calibration for Ultra-Wideband (UWB) networks, extending a Bayesian approach based on grid-based uncertainty propagation. The enhanced algorithm reduces measurement availability requirements while maintaining positioning accuracy through probabilistic state estimation. We validate the approach using real-world data from controlled indoor experiments with 12 nodes in a static environment. Experimental evaluation yields 0.28~m mean ranging error under line-of-sight conditions and 1.11~m overall ranging error across mixed propagation scenarios. Results confirm the algorithm's resilience to measurement noise and partial connectivity scenarios typical in industrial deployments. We evaluate algorithm robustness under NLOS-contaminated initialization, showing graceful accuracy degradation (median error 0.62--0.99~m) compared to closed-form methods that exhibit substantial performance collapse (median error up to 2.43~m). The findings contribute to automated UWB network initialization for indoor positioning applications, reducing infrastructure dependency compared to manual anchor calibration procedures.
\end{abstract}

\begin{IEEEkeywords}
Auto-positioning, Markov Localization, Ultra-Wideband (UWB), Radio-based Localization, Wireless Sensor Network (WSN), Bayes Filter
\end{IEEEkeywords}

\section{Introduction}
\IEEEPARstart{L}{ocation-aware} systems in intelligent transportation, smart cities, and logistics require sub-decimeter positioning accuracy for autonomous navigation, asset tracking, and human-machine interaction \cite{hailu2024theories}. Traditional radio-based localization relies on stationary anchors with precisely surveyed positions, creating deployment complexity that scales superlinearly with network density. Manual anchor calibration represents a substantial percentage of system installation costs \cite{herbruggen2023multihop}. Collaborative networks intensify this challenge through systematic error propagation. Positioning errors create correlated uncertainties that compound throughout the system.

\begin{figure}[!t]
\centerline{\includegraphics[width=1\columnwidth, trim= 0cm 0cm 0cm 0cm, clip]{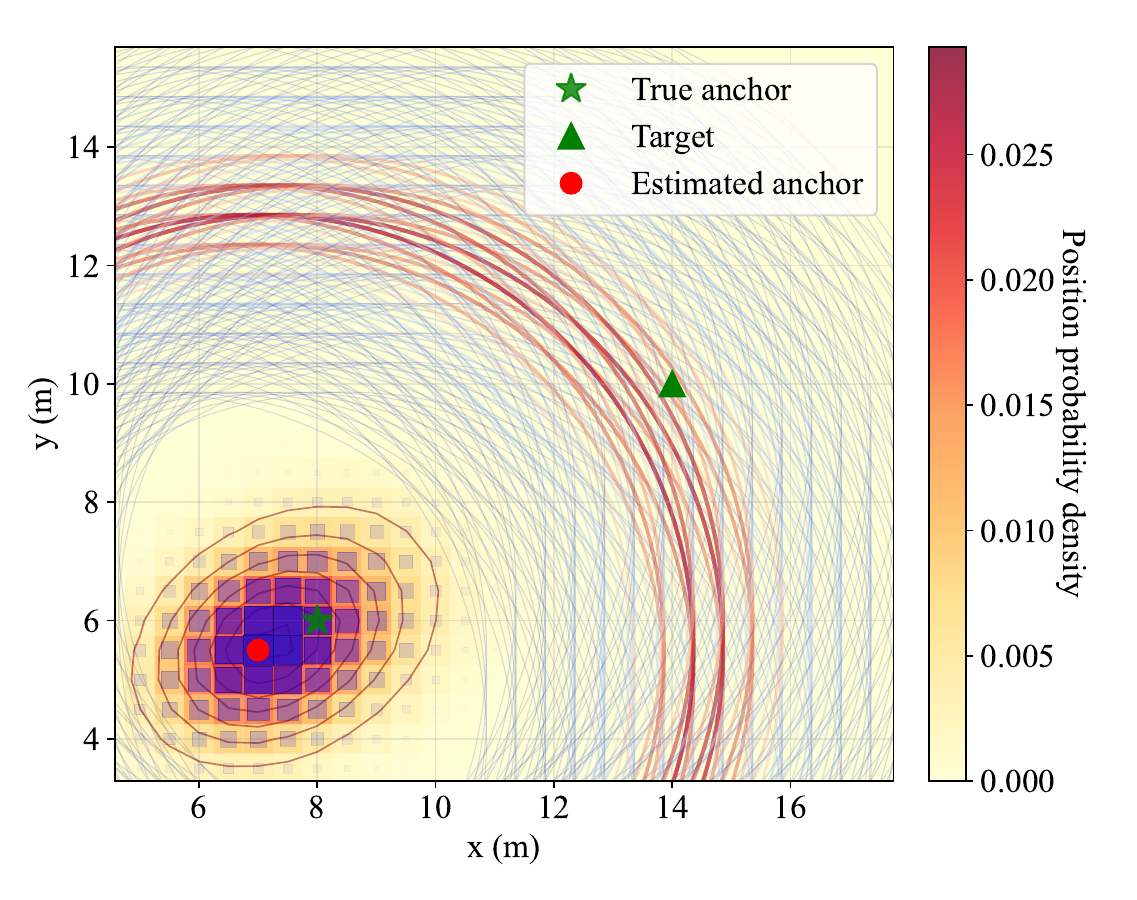}}
\caption{Grid-based uncertainty propagation for collaborative self-calibration: Anchor probability distribution generates multiple weighted measurement hypotheses (radiating lines) rather than single point-estimate constraints, preserving spatial uncertainty information for reliable localization.}
\label{fig:autopos}
\end{figure}

Collaborative self-calibration addresses these limitations by estimating node positions without a-priori knowledge of network topology \cite{hossain2015survey}. However, parametric methods exhibit critical limitations when addressing non-Gaussian distributions and multimodal uncertainties prevalent in non-line-of-sight (NLOS) environments \cite{herbruggen2023multihop}. Recent surveys identify inadequate noise models and insufficient uncertainty quantification as fundamental failure modes \cite{Kordi2024SurveyIPSDeepLearning, hailu2024theories}.

This research addresses collaborative self-calibration through probabilistic uncertainty propagation frameworks that achieve sub-meter positioning accuracy while reducing infrastructure dependencies. As illustrated in \cref{fig:autopos}, the proposed grid-based approach generates multiple weighted measurement hypotheses rather than single point-estimate constraints, preserving spatial uncertainty information for robust target localization. The methodology extends discrete Bayesian estimation to maintain uncertainty distributions across network nodes, validated through experimental evaluation using real-world data from a fully meshed UWB sensor network under mixed propagation conditions. 

This approach shows greater resilience to initialization parameters and measurement availability constraints compared to conventional parametric methods.

\subsection{Related Work}
\label{subsec:related_work}

Collaborative self-calibration encompasses automatic positioning procedures that enable 
anchor nodes to determine their locations without manual surveying, utilizing peer-to-peer 
ranging measurements \cite{ridolfi_self-calibration_2021, Schwarzbach2026Terminology_Survey}. This approach differs from 
cooperative localization, which focuses on information sharing to improve individual node 
accuracy \cite{wymeersch_cooperative_2009}, by eliminating fixed infrastructure dependencies 
entirely. The fundamental challenge emerges from uncertainty propagation characteristics 
in collaborative networks, where positioning errors systematically bias dependent node estimates.

\subsubsection{Collaborative calibration}
methods enhance accuracy for anchors with approximate known 
coordinates. Qi et al. \cite{qi_calibration_2024} address second-order error terms in 
trilateral localization through tightly coupled UWB-MEMS integration, demonstrating 
that anchor position errors significantly impact localization performance when neglected 
in conventional first-order approaches. Xu et al. \cite{xu_hybrid_calib_2020} employ 
Modified Light Gradient Boosting Models with cooperative screening strategies to handle 
estimation errors across diverse network topologies.

\subsubsection{Meshed anchor-tag networks}
Technology-specific implementations demonstrate meshed approaches across wireless technologies. 
WiFi-RTT systems \cite{raja_wifi-rtt_2024} utilize round-trip time measurements with SLAM 
approaches, while 5G-based systems \cite{lu_bayesian_5G_2023} leverage combined range and 
angle-of-arrival measurements. UWB-aided vision navigation \cite{hu_robust_vision_UAV_2023} 
and Calibration Unit approaches \cite{van_de_velde_CU_2013} provide alternative methodologies 
for inter-anchor distance estimation.

\subsubsection{Fully meshed networks}
achieve network-wide positioning without external references, 
representing the most autonomous positioning methodologies. Multidimensional Scaling (MDS) 
approaches form a foundational category, with Krapez et al. \cite{krapez_anchor_mds_2020} 
demonstrating MDS integration achieving 0.32~m anchor localization accuracy in three-dimensional 
environments. Closed-form auto-positioning approaches \cite{hamer_self-calibrating_baseline_2018,loahavilai_rapid_baseline_2021} 
provide computational efficiency through direct algebraic solutions, while machine learning 
enhanced methods \cite{ridolfi_uwb_ML_2021} achieve centimeter-level accuracy through 
adaptive physical layer settings and signal correction algorithms.

However, existing parametric collaborative localization methods exhibit significant 
limitations when addressing non-Gaussian distributions and multimodal uncertainties 
prevalent in NLOS propagation environments \cite{herbruggen2023multihop}. Traditional 
approaches approximate node position uncertainty through covariance matrices and 
trace operations \cite{Medina_hybrid_estimators_2020}, effectively discarding spatial 
distribution structure. This parametric representation fails to capture true network-wide uncertainty 
propagation, motivating non-parametric approaches that preserve discrete probability mass 
distributions throughout collaborative positioning processes.

\subsection{Scope and Contributions}
\label{subsec:scope}
This paper addresses the challenges of collaborative self-calibration in indoor positioning 
systems by evaluating a grid-based approach in a mixed LOS- and NLOS-environment, where 
traditional parametric methods exhibit diminished performance due to NLOS propagation and 
measurement outliers. The primary contributions of this work are fourfold: 

\begin{enumerate}
    \item We present a self-surveyed dataset acquired from a fully meshed UWB sensor network 
    employing two-way ranging (TWR), providing empirical validation data for ranging quality 
    assessment that captures challenges inherent to NLOS reception and intermittent signal loss.
    
    \item We extend and evaluate a previously introduced grid-based Bayesian formulation for 
    collaborative self-calibration \cite{jung_auto-positioning_2022}. This approach employs 
    non-parametric filtering that preserves uncertainty distributions across network nodes, 
    showing greater resilience compared to parametric self-localization methods.
    
    \item We show that the proposed method reduces connectivity requirements and
    measurement availability constraints compared to state-of-the-art approaches, enabling
    practical deployment in scenarios where conventional methods fail due to insufficient
    measurement density or network connectivity limitations.

    \item We evaluate algorithm robustness under NLOS-contaminated anchor initialization. The grid-based method maintains sub-meter median accuracy while closed-form approaches exhibit order-of-magnitude degradation, confirming practical applicability in environments where LOS initialization cannot be guaranteed.
\end{enumerate}

The remainder of this paper is structured as follows: \cref{sec:fundamentals} establishes 
theoretical foundations for collaborative self-calibration, introducing a baseline closed-form 
approach. \cref{sec:bayes} presents the proposed Bayesian grid-based approach with uncertainty 
propagation. \cref{sec:exp} describes the measurement campaign and UWB ranging data collection. 
\cref{sec:valid} presents empirical results with comparative positioning performance analysis. 
The paper concludes with a synthesis of findings in \cref{sec:conclusion}.

\section{Fundamentals}
\label{sec:fundamentals}

This section establishes the theoretical foundation for collaborative self-calibration by examining dependency structures in network-based positioning systems and presenting both baseline and proposed approaches. The analysis reveals limitations of conventional parametric uncertainty propagation methods, motivating the main contribution of this research.

\subsection{Collaborative Positioning Dependencies}
\label{subsec:dependencies}

\begin{figure}[htbp]
\includegraphics[width=1\columnwidth, trim= 0cm 0cm 0cm 0cm, clip]{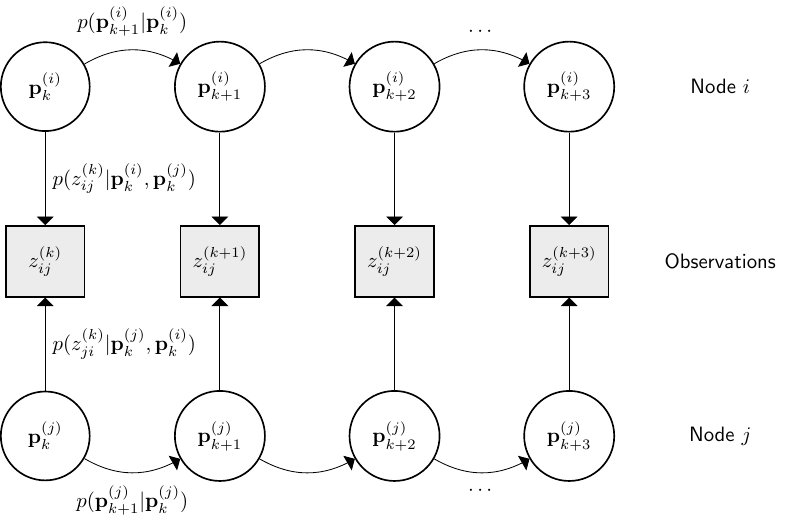}
\caption{Dependencies in collaborative positioning networks. Circles denote node states $\mathbf{p}_k^{(i)}$, squares represent observations $z_{ij}^{(k)}$, horizontal arrows indicate temporal evolution $p(\mathbf{p}_{k+1}^{(i)}|\mathbf{p}_k^{(i)})$, and vertical arrows represent measurement dependencies $p(z_{ij}^{(k)}|\mathbf{p}_k^{(i)}, \mathbf{p}_k^{(j)})$. This coupling structure creates correlated uncertainties that propagate throughout the network.}
\label{fig:collaborative_dependencies}
\end{figure}

Collaborative positioning systems exhibit interdependencies that distinguish them from conventional anchor-based localization. Figure~\ref{fig:collaborative_dependencies} illustrates the fundamental dependency structure, where $\mathbf{p}_k^{(i)}$ denotes the position vector of node $i$ at time step $k$, $z_{ij}^{(k)}$ represents the distance measurement between nodes $i$ and $j$, and $p(\cdot|\cdot)$ denotes conditional probability distributions.

Temporal evolution follows prediction models $p(\mathbf{p}_{k+1}^{(i)}|\mathbf{p}_k^{(i)})$ representing positional changes between time steps. The critical challenge emerges from spatial coupling: distance measurements $z_{ij}^{(k)}$ depend simultaneously on both node positions through $p(z_{ij}^{(k)}|\mathbf{p}_k^{(i)}, \mathbf{p}_k^{(j)})$. This bidirectional dependency creates systematic bias propagation where positioning errors in individual nodes affect estimates of dependent nodes.

Traditional parametric approaches approximate position uncertainties as additive noise terms, severing the dependency links in Figure~\ref{fig:collaborative_dependencies}. This approximation discards distributional information, particularly problematic for non-Gaussian or multimodal uncertainties common in NLOS environments. The resulting estimates become overconfident, failing to capture network-wide uncertainty propagation.

Collaborative systems require estimation methods that preserve and propagate uncertainty information across network nodes while maintaining computational tractability.

\subsection{Baseline Closed-Form Approach}
\label{subsec:closed-form}

\begin{figure}[htbp]
\centerline{\includegraphics[width=.7\columnwidth]{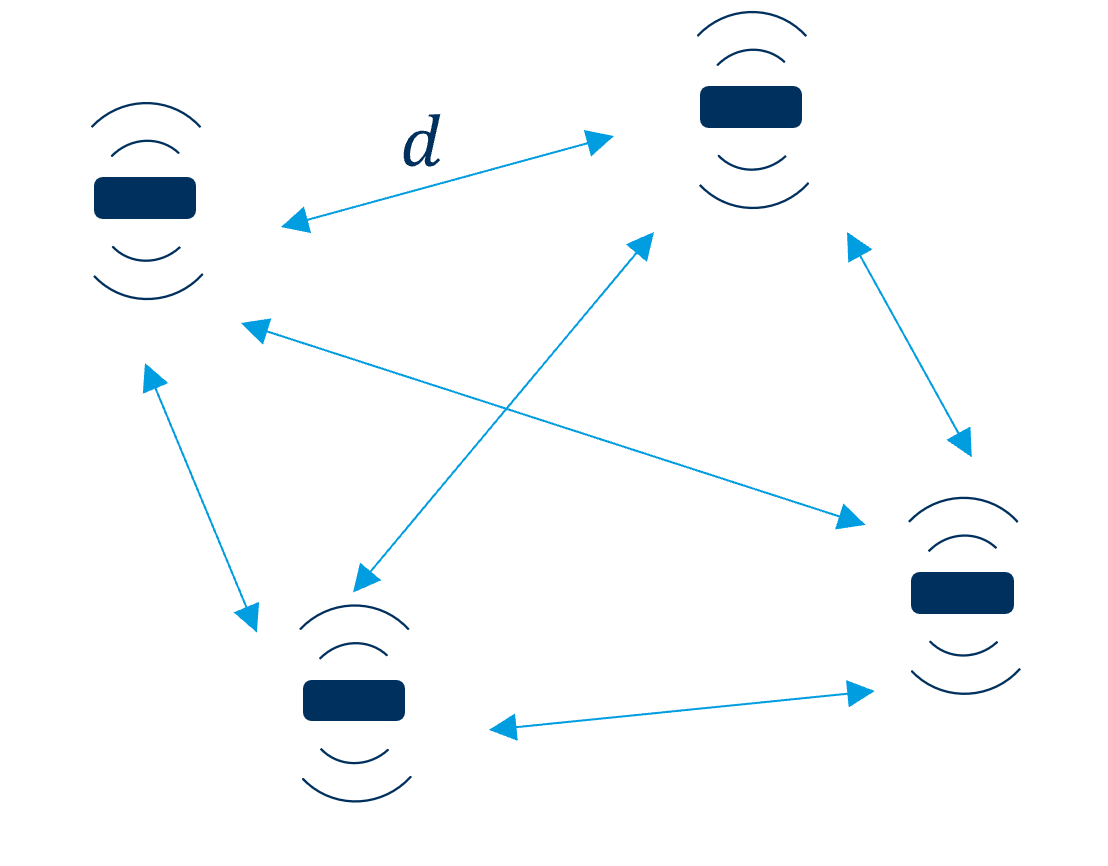}}
\caption{Fully meshed sensor network configuration}
\label{fig:network}
\end{figure}

The baseline method, utilizing a fully meshed network as depicted in \cref{fig:network}, establishes a coordinate reference frame using three anchors with the following assumptions \cite{hamer_self-calibrating_baseline_2018,loahavilai_rapid_baseline_2021}:

\begin{itemize}
\item $\mathbf{a}_0$ is situated at the coordinate origin
\item The direction from $\mathbf{a}_0$ to $\mathbf{a}_1$ defines the positive x-axis  
\item $\mathbf{a}_2$ lies in the half-plane with positive y-coordinate
\end{itemize}

The objective is estimating two-dimensional positions $\mathbf{p}_i = [x_i, y_i]^T$ for each node $i \in \{1, ..., N\}$ using the inter-node distance matrix:

\begin{equation}
    \mathbf{D}^{(k)} = 
    \begin{pmatrix}
    z_{0,0}^{(k)} & z_{0,1}^{(k)} & \cdots & z_{0,N}^{(k)} \\
    z_{1,0}^{(k)} & z_{1,1}^{(k)} & \cdots & z_{1,N}^{(k)} \\
    \vdots  & \vdots  & \ddots & \vdots  \\
    z_{N,0}^{(k)} & z_{N,1}^{(k)} & \cdots & z_{N,N}^{(k)}
    \end{pmatrix}
    \label{eq:distmatrix}
\end{equation}

where $z_{ij}^{(k)}$ represents measured distance between nodes $i$ and $j$ at time step $k$. We distinguish between true geometric distances $d_{ij} = \|\mathbf{p}_i - \mathbf{p}_j\|_2$ and measured distances $z_{ij}^{(k)} = d_{ij} + n_{ij}^{(k)}$, where $n_{ij}^{(k)}$ represents measurement error.

The distance matrix is populated through direct pairwise ranging measurements between all node combinations. Our fully meshed architecture performs direct two-way ranging (TWR) between each node pair without intermediate relay nodes, distinguishing this approach from multi-hop collaborative localization methods \cite{herbruggen2023multihop}. While the experimental deployment provides a fully meshed network, the proposed grid-based method does not require a complete distance matrix. The probabilistic framework accommodates partial connectivity through selective reference anchor utilization in the measurement update, as demonstrated in Section~\ref{subsec:availability}.

The coordinate frame is established with $\mathbf{a}_0 = [0, 0]^T$ and $\mathbf{a}_1 = [z_{01}^{(k)}, 0]^T$. Subsequent nodes are positioned by:

\begin{equation}
    \mathbf{a}_2 = \left[ \frac{(z_{02}^{(k)})^2 - (z_{12}^{(k)})^2 + x_{1}^2}{2x_{1}}, 
\sqrt{(z_{02}^{(k)})^2 - x_{2}^2}\right]^T \; .
\label{eq:Anchor2}
\end{equation}

This approach exhibits several limitations that motivate the probabilistic framework presented in Section~\ref{sec:bayes}: (1) measurement errors propagate algebraically through squared terms without uncertainty quantification, (2) measurement failures or high ranging errors render trilateration infeasible due to geometric inconsistencies, (3) the square root operation introduces numerical instabilities when $d_{02}^2 < x_2^2$, and (4) the deterministic coordinate frame depends critically on the first three anchor positions, creating systematic sensitivity to initialization selection as demonstrated in Section~\ref{subsec:positioning}.

\section{Bayesian Grid-Based Approach}
\label{sec:bayes}
The key methodological innovation extends the grid-based framework of \cite{jung_auto-positioning_2022} by implementing hypothesis-based uncertainty propagation rather than parametric approximation of anchor position uncertainties. While parametric approaches collapse probability distributions to covariance matrices, the proposed method preserves and propagates discrete probability mass from anchors, enabling more accurate uncertainty quantification in collaborative positioning scenarios. 

\begin{figure}[htbp]
    \centering
    \includegraphics[width=.9\linewidth]{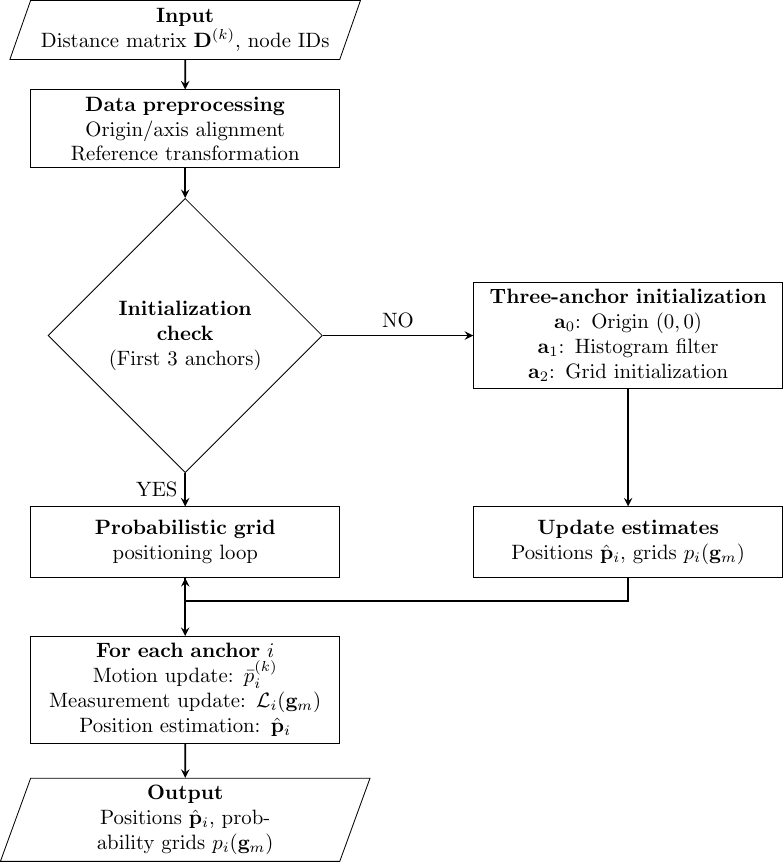}
    \caption{Algorithmic flowchart of the grid-based self-calibration method. The process implements distance matrix preprocessing, sequential three-anchor initialization ($\mathbf{a}_0$ at origin, $\mathbf{a}_1$ via histogram-, $\mathbf{a}_2$ via grid filter), and recursive Bayesian filtering with uncertainty propagation.}
    \label{fig:flowchart}
\end{figure}

This is achieved by incorporating weighted contributions from multiple position hypotheses into the likelihood calculation as part of the correction step in a recursive state estimation framework. The overall procedure is summarized in \cref{fig:flowchart} and detailed in the following.

Discrete grid representation of the state space enables non-parametric probability density estimation that is resilient to non-Gaussian measurement noise and multimodal distributions. For each node, the continuous position space is discretized into $M$ equidistant grid points $\mathcal{G} = \{\mathbf{g}_1, \mathbf{g}_2, \ldots, \mathbf{g}_M\}$.

Posterior probabilities over the grid are maintained and updated following Bayes' rule:

\begin{equation}
\text{posterior} \propto \text{likelihood} \times \text{prior}
\end{equation}

\subsubsection{Parameter Selection}
\label{sssec:parameter_selection}

The grid-based approach employs the following parameters:
\begin{itemize}
\item $M$ -- Total number of discretized grid cells
\item $\tau$ -- Percentage threshold for probability mass selection
\item $r_{\text{est}}$ -- Estimation region radius  
\item $\sigma_r$ -- Ranging measurement standard deviation
\item $v$ -- Expected node velocity 
\item $\sigma_v$ -- Velocity uncertainty standard deviation
\end{itemize}

Table~\ref{tab:parameters} summarizes parameter values and selection rationale. Grid resolution balances spatial accuracy against computational cost, whereas the probability threshold $\tau$ captures primary hypotheses while limiting computation. Ranging noise $\sigma_r$ derives from LOS statistics (Table~\ref{tab:ranging_quality}) with safety margin.

\begin{table}[htbp]
\centering
\caption{Grid-based method parameters.}
\label{tab:parameters}
\begin{tabular}{@{}lll@{}}
\toprule
Parameter & Value & Rationale \\
\midrule
$M$ & $\sim$13,200 & 0.1\,m spacing over 44$\times$30\,m\textsuperscript{2} \\
$\tau$ & 3\% & Primary hypotheses \\
$r_{\text{est}}$ & 2\,m & Local averaging without over-smoothing \\
$\sigma_r$ & 0.5\,m & LOS statistics + margin \\
$\sigma_v$ & 0.1\,m & Filter recovery in static scenarios \\
\bottomrule
\end{tabular}
\end{table}

\subsection{Initialization Phase}

The proposed method leverages the coordinate frame established in \cref{subsec:closed-form} while implementing probabilistic estimation. Unlike the deterministic closed-form approach, the initialization phase maintains uncertainty distributions throughout coordinate system establishment.

\subsubsection{Anchor 0 - Origin Definition}
Following the baseline methodology, the first anchor defines the coordinate system origin $\mathbf{a}_0 = [0, 0]^T$. No uncertainty is associated with this mandatory definition.

\subsubsection{Anchor 1 - Histogram Filter}
The second anchor position is estimated along the positive x-axis using a one-dimensional histogram filter with $H$ hypotheses over the range $[0, d_{\text{max}}]$: 

\begin{equation}
p_1(h_j) = \frac{\exp\left(-\frac{(z_{01}^{(k)}-h_j)^2}{2\sigma_r^2}\right)}{\sqrt{2\pi\sigma_r^2}} \; ,
\label{eq:histogram_filter}
\end{equation}
where $h_j$ represent the histogram bins, $\sigma_r$ denotes the ranging measurement standard deviation, and $d_{\text{max}}$ represents the maximum operational range. The position estimation reduces to a one-dimensional problem while maintaining probabilistic uncertainty representation. The non-parametric nature of the histogram filter enables incorporation of arbitrary probability distributions without restrictive Gaussian assumptions, facilitating stable estimation under multimodal or heavy-tailed measurement error conditions.

\subsubsection{Anchor 2 - Grid Initialization}

The third anchor represents the first implementation of the grid-based uncertainty propagation framework, establishing the foundational probabilistic estimation method applied to all subsequent anchors. Unlike the deterministic origin definition $\mathbf{a}_0$ and one-dimensional histogram approach $\mathbf{a}_1$, $\mathbf{a}_2$ employs full two-dimensional Bayesian grid estimation. The initialization procedure utilizes three inter-node distance measurements $(z_{01}^{(k)}, z_{02}^{(k)}, z_{12}^{(k)})$ within the grid-based measurement update framework. The trilateration constraints from \cref{subsec:closed-form} provide geometric initialization of the applicable state space. The grid-based uncertainty propagation processes the three distance measurements through the complete Bayesian framework:

\begin{equation}
p_2(\mathbf{g}_m) = \eta \cdot \prod_{j \in \{0,1\}} \exp\left(-\frac{(\|\mathbf{g}_m - \mathbf{a}_j\| - z_{j2}^{(k)})^2}{2\sigma_r^2}\right)
\label{eq:a2_grid_update}
\end{equation}

where $\eta$ is the normalization constant ensuring $\sum_m p(\mathbf{g}_m) = 1$ and $\mathbf{a}_j$ denotes the established positions of anchors 0 and 1. The subsequent position estimation $\hat{\mathbf{a}}_2$ follows the approach detailed in \cref{ssec:position_estimation}.  

\subsubsection{Initialization Robustness}
The probabilistic initialization provides inherent robustness to NLOS-contaminated measurements compared to closed-form trilateration. Rather than computing a single point solution that fails under geometric inconsistency, Equation~(\ref{eq:a2_grid_update}) weights grid cells by measurement consistency across all three initialization distances. NLOS bias in one measurement creates probability mass in regions potentially inconsistent with the remaining measurements, resulting in broader but more representative uncertainty distributions.

To evaluate initialization robustness, we define two additional configurations (Conf.~3 and Conf.~4) where 2 of 3 initial anchor links are NLOS-contaminated (see \cref{subsec:positioning}). The grid-based method achieves median errors of 0.62--0.99\,m under NLOS initialization compared to 0.34--0.68\,m under LOS, demonstrating graceful degradation. In contrast, the closed-form approach exhibits median errors of 1.48--2.43\,m with extreme outliers exceeding 30\,m, confirming the robustness advantage of probabilistic uncertainty propagation.

\subsection{Recursive Position Estimation}

\begin{figure}[ht]
    \centering
    \includegraphics[width=\linewidth]{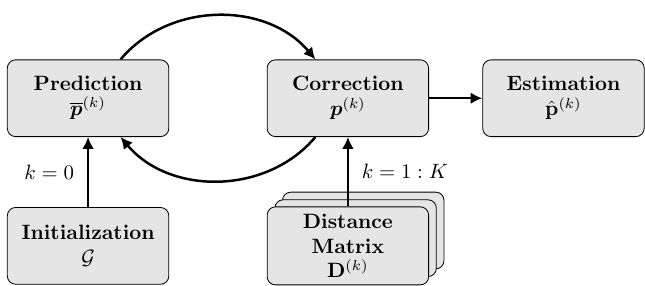}
    \caption{Recursive filter structure for grid-based self-calibration.}
    \label{fig:cgp_structure}
\end{figure}

After initialization, the remaining anchors employ a recursive Bayesian filter structure (\cref{fig:cgp_structure}), alternating between prediction and update steps.

\begin{figure*}[htbp!]
    \centering
    \begin{subfigure}[b]{0.275\linewidth}
    \centering
    \includegraphics[width=\linewidth, trim= 0cm 0cm 0cm 0cm, clip]{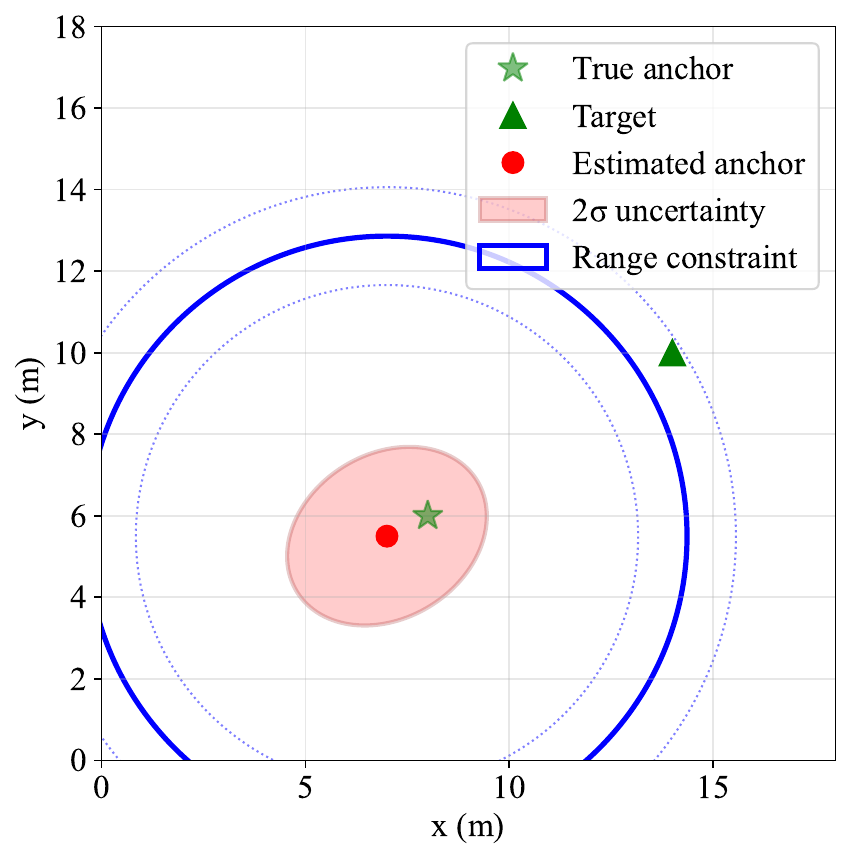}
    \caption{}
    \label{fig:fig1_parametric}
    \end{subfigure}
    \centering
    \begin{subfigure}[b]{0.325\linewidth}
    \centering
    \includegraphics[width=\linewidth, trim= 0cm 0cm 0cm 0cm, clip]{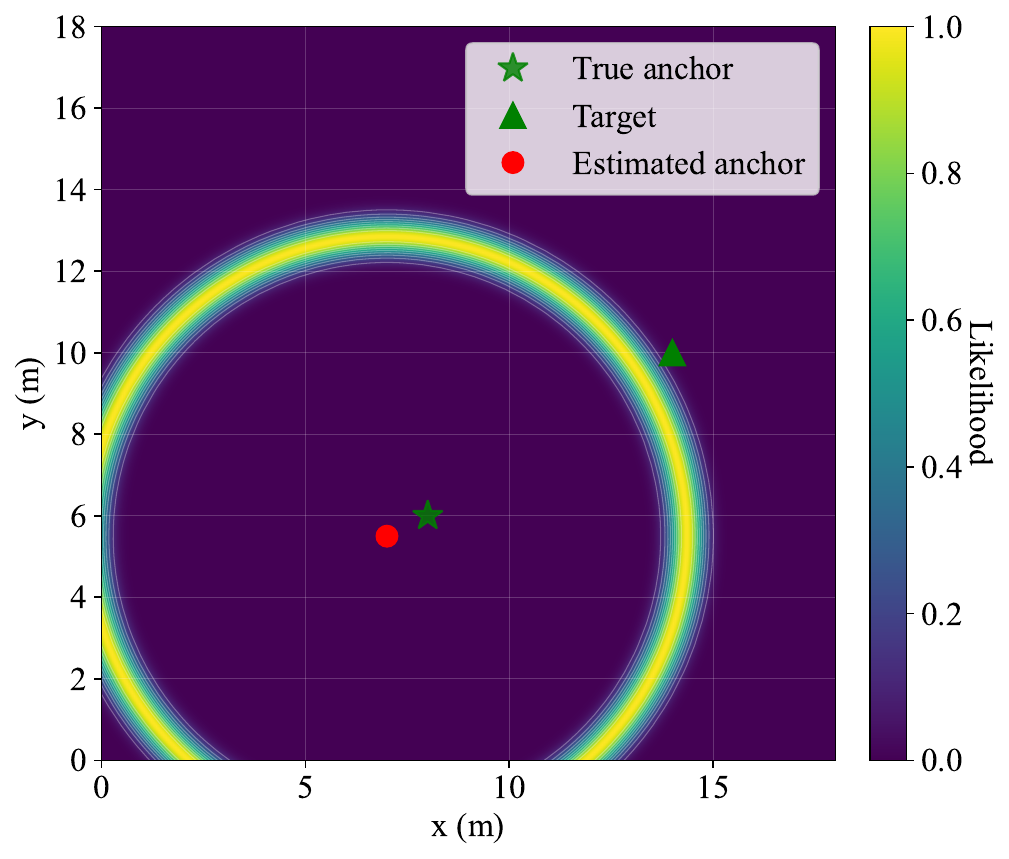}
    \caption{}
    \label{fig:fig3_parametric_likelihood}
    \end{subfigure}
    \centering
    \centering
    \begin{subfigure}[b]{0.32\linewidth}
    \centering
    \includegraphics[width=\linewidth, trim= 0cm 0cm 0cm 0cm, clip]{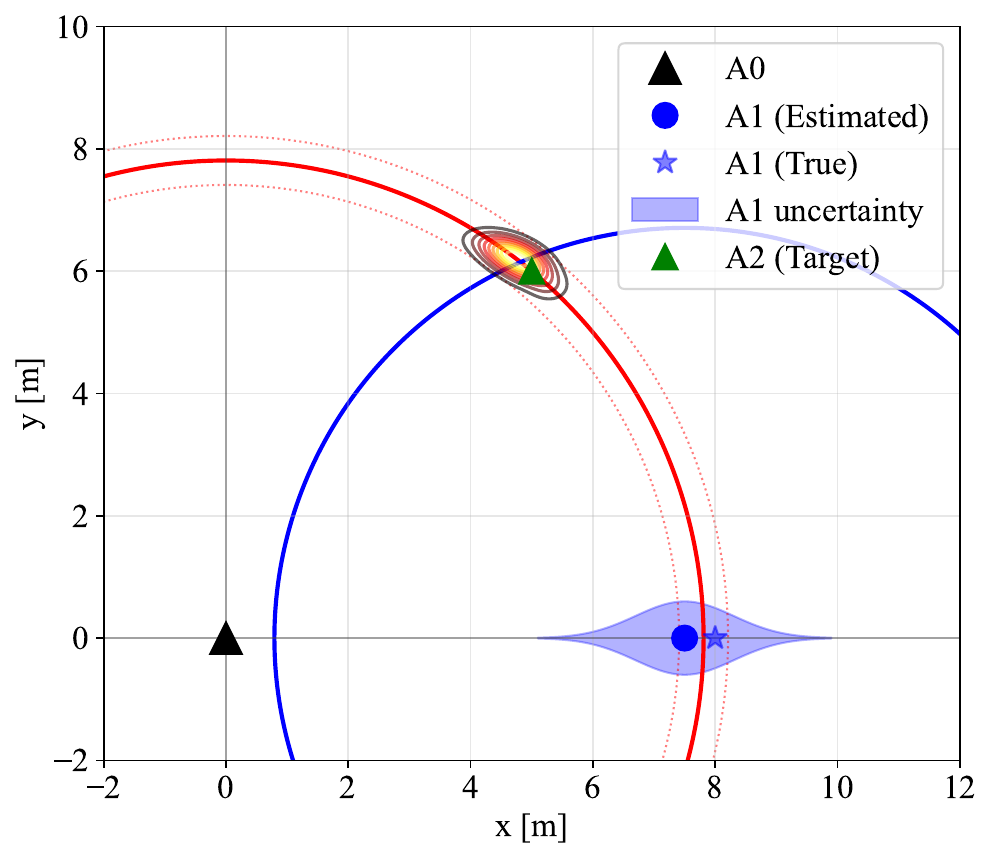}
    \caption{}
    \label{fig:fig_init_1_parametric}
    \end{subfigure}
    \centering
    \begin{subfigure}[b]{0.275\textwidth}
    \centering
    \includegraphics[width=\linewidth, trim= 0cm 0cm 0cm 0cm, clip]{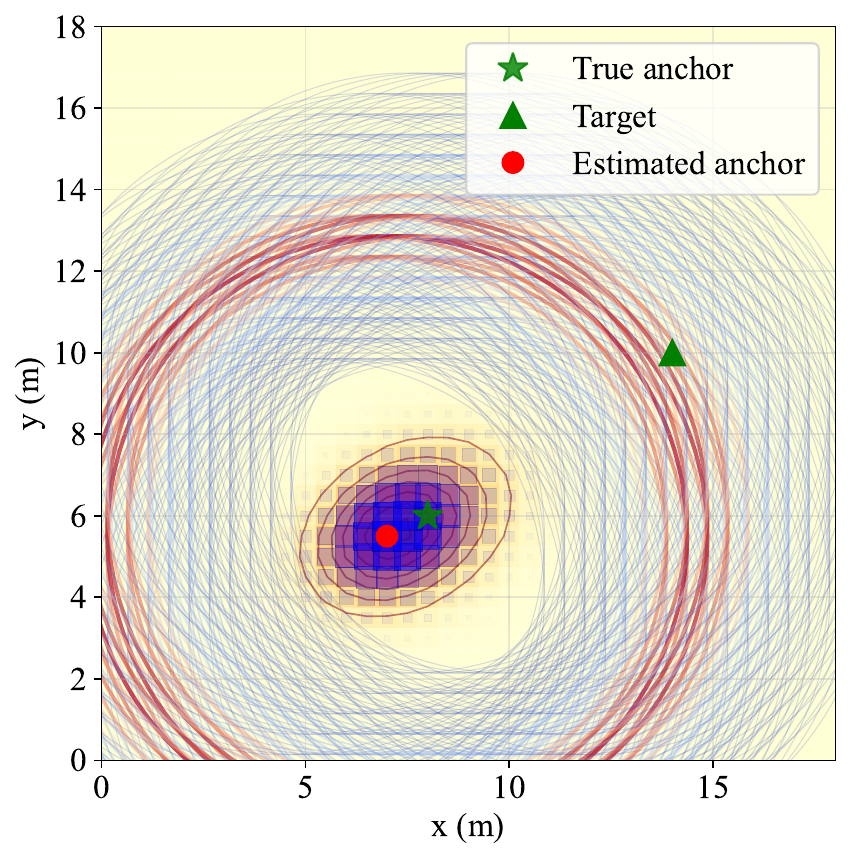}
    \caption{}
    \label{fig:fig2_grid_based}
    \end{subfigure}
    \centering
    \begin{subfigure}[b]{0.325\linewidth}
    \centering
    \includegraphics[width=\linewidth, trim= 0cm 0cm 0cm 0cm, clip]{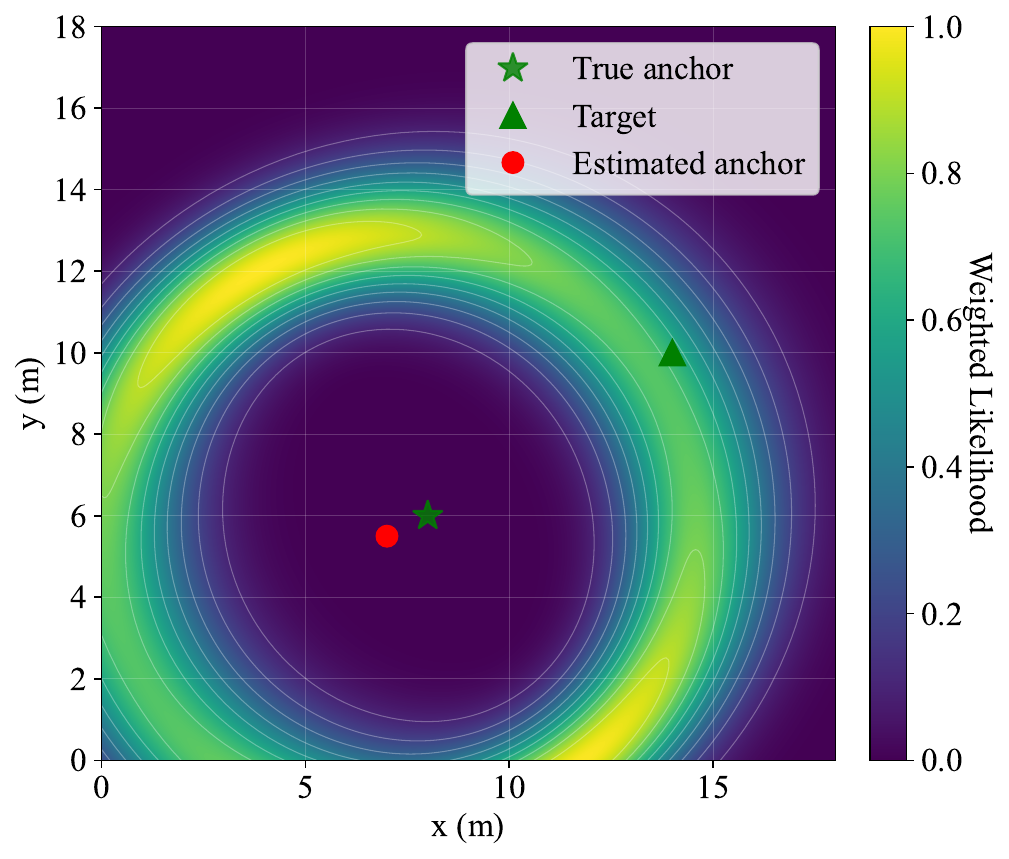}
    \caption{}
    \label{fig:fig4_grid_likelihood} 
    \end{subfigure}
    \centering
    \begin{subfigure}[b]{0.32\linewidth}
    \centering
    \includegraphics[width=\linewidth, trim= 0cm 0cm 0cm 0cm, clip]{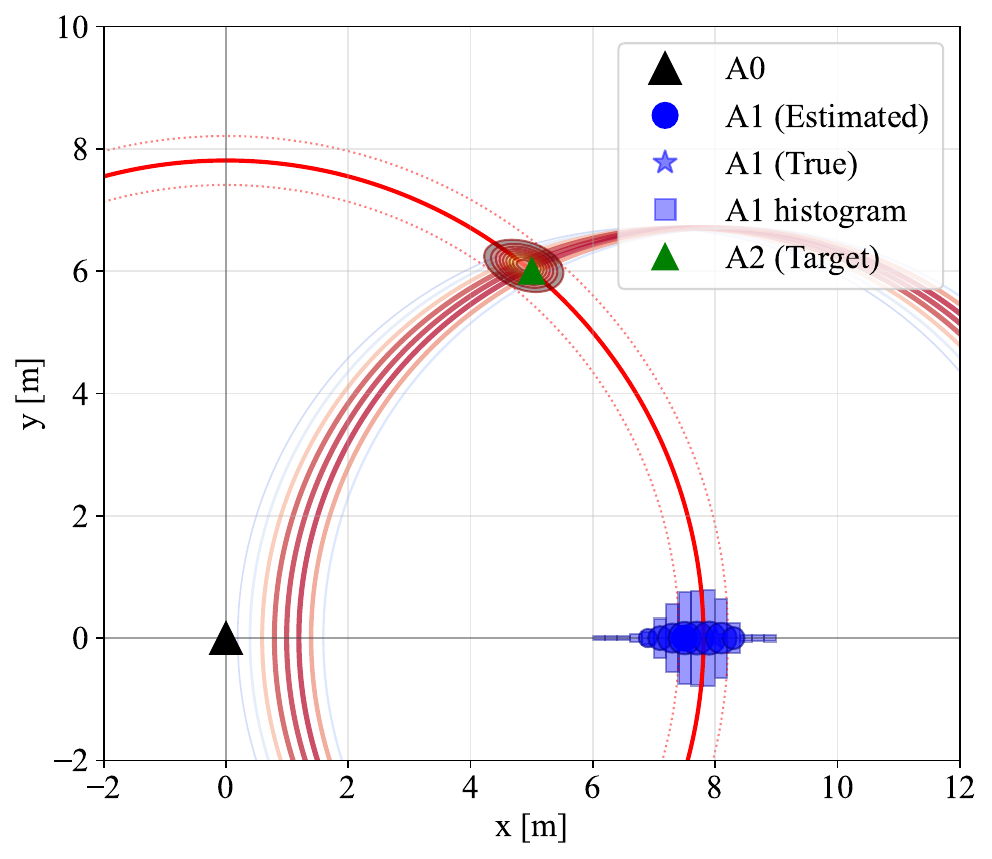}
    \caption{}
    \label{fig:fig_init_2_nonparametric}
    \end{subfigure}
    \centering
    \caption{Uncertainty propagation for collaborative self-calibration: Top row presents traditional point-estimate approach; bottom row demonstrates the proposed grid-based method. \textbf{(a)} Traditional parametric representation of reference anchor as point estimate with associated covariance. \textbf{(d)} Grid-based representation showing probability distribution (colormap), sampled position hypotheses (colored points), and measurement constraints (circles). \textbf{(b, e)} Likelihood computation: (b) single measurement constraint from point estimate; (e) weighted contributions from multiple position hypotheses demonstrating aggregate likelihood fields. \textbf{(c)} Parametric initialization procedure. \textbf{(f)} Grid-based initialization using hypotheses-based uncertainty propagation.}
    \label{fig:uncertainty_propagation}
\end{figure*}

\subsubsection{Motion update}
The prediction step incorporates expected node movement:

\begin{equation}
\bar{p}_i^{(k)} = \sum_{m \in \mathcal{K}} p_m^{(k-1)} \cdot p(\mathbf{g}_i^{(k)} | \mathbf{g}_m^{(k-1)})
\end{equation}

where $\mathcal{K}$ represents the set of grid cells with the highest probabilities. This selection reduces computational complexity while maintaining estimation accuracy. The transition probability is modeled as:

\begin{equation}
p(\mathbf{g}_i^{(k)} | \mathbf{g}_m^{(k-1)}) = \mathcal{N}(\|\mathbf{g}_i - \mathbf{g}_m\|_2 - v\Delta t, \sigma_v^2)
\end{equation}

with velocity $v$ and velocity uncertainty $\sigma_v$. For static scenarios, $v = 0$ but $\sigma_v > 0$ adds noise to enable filter recovery from erroneous estimates.

\subsubsection{Measurement update with uncertainty propagation}
\label{ssec:uncertainty_propagation}

The core contribution of this measurement update addresses uncertainty propagation in collaborative positioning through discrete probability mass preservation rather than parametric approximation. Traditional approaches approximate reference position uncertainty through covariance matrices and apply trace operations \cite{Medina_hybrid_estimators_2020}:

\begin{equation}
\sigma_{1,2} = \sigma_r + \sqrt{\text{tr}(\mathbf{P}_1)} + \sqrt{\text{tr}(\mathbf{P}_2)}
\label{eq:parametric_uncertainty}
\end{equation}

to compute aggregate measurement noise. This parametric representation discards spatial distribution, particularly problematic when positions exhibit multimodal or non-Gaussian distributions common in NLOS-contaminated environments.

The proposed hypotheses-based approach preserves distributional information by maintaining discrete probability mass representations. Rather than collapsing anchor $j$'s position distribution to a single covariance matrix, the method samples the top $\tau\%$ probability mass as discrete hypotheses:

\begin{equation}
\mathcal{T}_j^{\tau} = \{\mathbf{g}_{jl} : l \in \arg\max_m p_j(\mathbf{g}_m)\}
\label{eq:hypothesis_sampling}
\end{equation}

where each hypothesis retains its associated probability weight $w_{jl} = p_j(\mathbf{g}_{jl})/\sum_{k \in \mathcal{T}_j^{\tau}} p_j(\mathbf{g}_{jk})$.

For anchor $i$ being updated, the measurement likelihood incorporates uncertainty from all anchors $j \in \mathcal{N}_i$ through weighted contributions from multiple position hypotheses:

\begin{equation}
\mathcal{L}_i(\mathbf{g}_m) = \prod_{j \in \mathcal{N}_i} \sum_{l \in \mathcal{T}_j^{\tau}} w_{jl} \cdot \exp\left(-\frac{(\|\mathbf{g}_m - \mathbf{g}_{jl}\|_2 - z_{ij}^{(k)})^2}{2\sigma_r^2}\right)
\label{eq:likelihood_uncertainty}
\end{equation}

where $\mathcal{N}_i$ is the set of available reference anchors for node $i$, $\mathcal{T}_j^{\tau}$ represents the top $\tau\%$ probability cells of anchor $j$, $w_{jl}$ are the normalized weights for each hypothesis and $z_{ij}^{(k)}$ is the measured distance between nodes $i$ and $j$ at time step $k$.

This weighted superposition captures the true uncertainty propagation from anchors to target positions, as illustrated in \cref{fig:uncertainty_propagation}. The parametric approximation generates circular measurement constraints from point estimates, while the hypotheses-based method produces aggregate likelihood fields that reflect the actual spatial uncertainty distribution of anchor positions. This distinction becomes increasingly important as positioning errors accumulate through the network.

The posterior probability is updated using Bayes' rule with normalization constant $\eta$:

\begin{equation}
p_i(\mathbf{g}_m) = \eta \cdot \bar{p}_i(\mathbf{g}_m) \cdot \mathcal{L}_i(\mathbf{g}_m)
\label{eq:posterior_update}
\end{equation}

\subsubsection{Position estimation}
\label{ssec:position_estimation}

The position estimate at each time step is computed as a weighted average in the circular region of radius $r$ around the maximum likelihood estimate with $\mathcal{B}_r(\mathbf{g}_{\text{MLE}}) = \{m : \|\mathbf{g}_m - \mathbf{g}_{\text{MLE}}\|_2 < r\}$:

\begin{equation}
\hat{\mathbf{p}}_i = \frac{\sum_{m \in \mathcal{B}_r(\mathbf{g}_{\text{MLE}})} p_i(\mathbf{g}_m) \cdot \mathbf{g}_m}{\sum_{m \in \mathcal{B}_r(\mathbf{g}_{\text{MLE}})} p_i(\mathbf{g}_m)} \; .
\end{equation}

\subsubsection{Computational complexity}
\label{sssec:complexity}

The per-epoch complexity scales as $\mathcal{O}(\tau^2 M^2 N^2)$. For our configuration
with $M \approx 13{,}200$ grid cells, $N = 12$ nodes, and probability threshold
$\tau = 0.03$, this yields approximately $2.3 \times 10^7$ likelihood evaluations per
network update. The centralized architecture aggregates all $N(N-1)/2 = 66$ pairwise
measurements at a single coordinator. Future distributed implementations could reduce
per-node computation to $\mathcal{O}(\tau^2 M^2 |\mathcal{N}_i|)$ by restricting
likelihood updates to local neighbor sets.

\section{Experimental Setup and Data Collection}
\label{sec:exp}

To validate the proposed grid-based positioning method, we conducted controlled experiments in a real-world industrial environment within a business park in Torgau, Germany. The experimental campaign utilized a fully meshed UWB sensor network enabling comprehensive inter-node ranging measurements under mixed propagation conditions.

\subsection{Hardware Configuration}
\label{subsec:hardware}

The experimental setup employed 12 Qorvo MDEK1001 UWB transceivers operating in the 3.5-6.5~GHz frequency band. The original firmware was customized by Deveritec GmbH under proprietary licensing to enable full mesh networking capabilities, supporting peer-to-peer ranging between all node pairs through direct two-way ranging (TWR) without intermediate relay hops. Architectural descriptions of similar fully meshed UWB implementations using the same hardware platform are available in \cite{Jimnez2021MDEKFirmware}. 

One node serves as network coordinator, responsible for orchestrating the sequential ranging operations to prevent channel conflicts through time-division multiplexing. The network operated with 10~Hz measurement frequency, representing the aggregate system update rate after all pairwise ranging operations complete sequentially within each epoch. The centralized coordinator is also responsible for data aggregation.

\begin{figure*}[htbp!]
    \centering
    \begin{subfigure}[b]{0.24\linewidth}
        \centering
        \includegraphics[width=\linewidth]{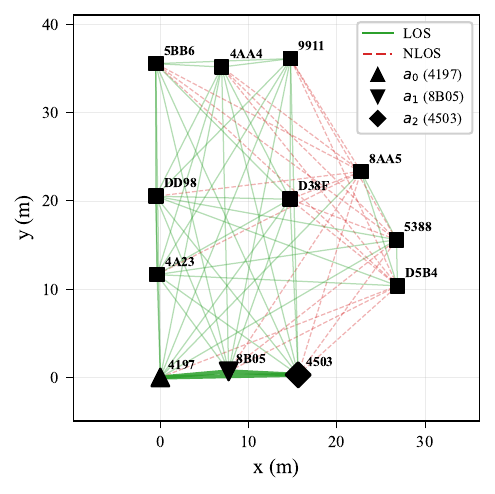}
        \caption{Conf.\ 1 (LOS)}
        \label{fig:arrange_conf1}
    \end{subfigure}
    \hfill
    \begin{subfigure}[b]{0.24\linewidth}
        \centering
        \includegraphics[width=\linewidth]{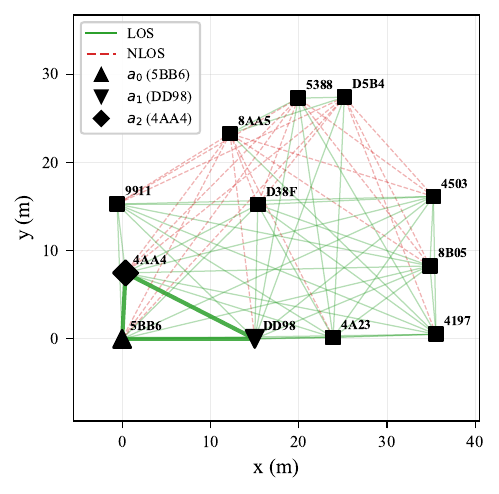}
        \caption{Conf.\ 2 (LOS)}
        \label{fig:arrange_conf2}
    \end{subfigure}
    \hfill
    \begin{subfigure}[b]{0.24\linewidth}
        \centering
        \includegraphics[width=\linewidth]{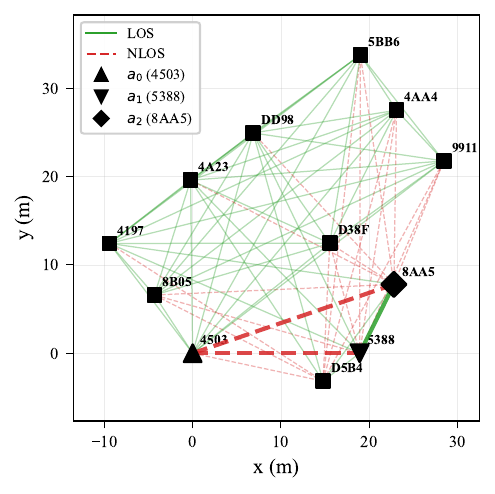}
        \caption{Conf.\ 3 (NLOS)}
        \label{fig:arrange_conf3}
    \end{subfigure}
    \hfill
    \begin{subfigure}[b]{0.24\linewidth}
        \centering
        \includegraphics[width=\linewidth]{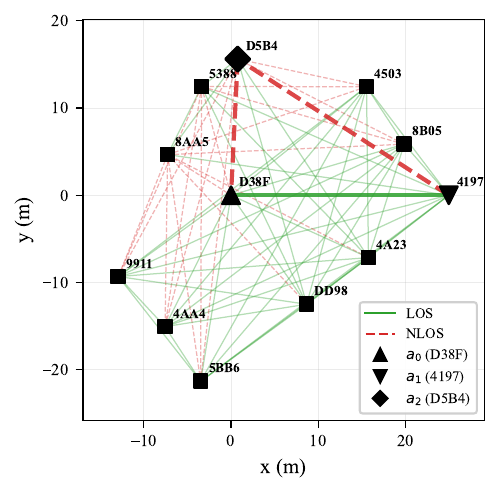}
        \caption{Conf.\ 4 (NLOS)}
        \label{fig:arrange_conf4}
    \end{subfigure}
    \caption{Node constellations and visibility conditions for four initialization configurations. Green solid lines indicate LOS links, red dashed lines indicate NLOS links. Thick lines highlight the initialization triangle ($\mathbf{a}_0$--$\mathbf{a}_1$--$\mathbf{a}_2$). (a,\,b)~Conf.~1 and~2: pure LOS initialization. (c,\,d)~Conf.~3 and~4: NLOS-contaminated initialization with 2/3 NLOS links in the initialization triangle.}
    \label{fig:arrange_conf}
\end{figure*}

\subsection{Deployment configuration and reference system}  
\label{subsec:deployment}

\begin{figure}[htbp!]
    \centering
    \includegraphics[width=.75\linewidth]{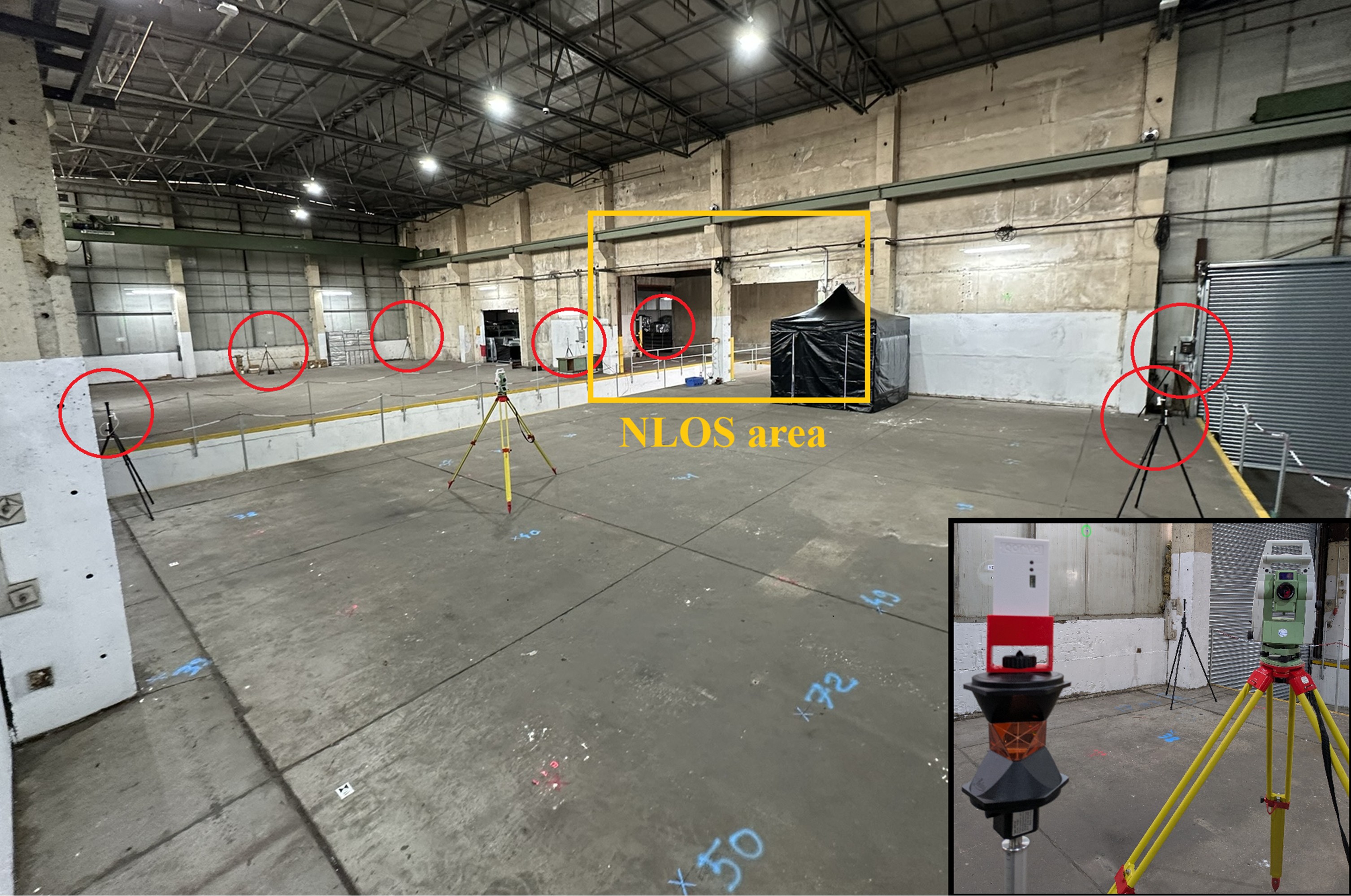}
    \caption{Measurement environment with distributed nodes (red) and total station setup for ground truth position surveying.}
    \vspace{-.7cm}
    \label{fig:realworld}
\end{figure}

We deployed a network of 12 nodes throughout a 44×30~m² indoor hall (see \cref{fig:realworld}), mounted on tripods at 1.60~m height. Node placement creates diverse propagation scenarios, including both LOS and NLOS configurations essential for robustness evaluation under realistic deployment conditions.

Nodes were powered using USB mains power supplies or USB battery packs, enabling flexible placement. The sensor network was configured and controlled via PC interface, providing access to sensor data and network parameters. 

We documented visibility conditions between node pairs, creating a binary LOS/NLOS classification matrix for subsequent propagation-dependent analysis. The resulting propagation conditions encompass moderate NLOS scenarios typical of industrial hall environments, characterized by partial occlusions from equipment and architectural elements rather than complete signal blockage.

We established reference coordinates using an optical Leica TS16 total station. A surveying prism served as the target for tachymeter measurements (see \cref{fig:realworld}). The total station recorded precise coordinates of the prism, enabling creation of a local coordinate system based on predefined reference points in the environment. 

For compatibility with the proposed algorithm, we applied a coordinate transformation. Two selected nodes define the positive x-axis, with the first node at the coordinate system origin, consistent with the assumptions in \cref{subsec:closed-form}. \Cref{fig:arrange_conf} illustrates the resulting four node configurations: two with pure LOS initialization and two with NLOS-contaminated initialization. This design enables systematic evaluation of anchor initialization sensitivity under varying propagation conditions.

\subsection{Dataset Characteristics}
\label{subsec:dataset}

The experimental campaign generated approximately $132,000$ individual ranging measurements across all node pairs over $2,000$ measurement epochs. Each epoch captured the complete $12 \times 12$ distance matrix with systematic documentation of ranging availability and residuals. Data are structured in text format with event timestamps, node identifiers, symmetric distance matrices, ground truth coordinates, and binary visibility condition matrices. Four anchor initialization configurations (\cref{fig:arrange_conf}), two with pure LOS and two with NLOS-contaminated initialization, enable systematic evaluation of algorithm sensitivity to reference frame selection and initialization propagation conditions. This supports reproducible performance assessment across varying measurement availability and propagation scenarios.

\section{Validation and Results}
\label{sec:valid}

This section validates the proposed grid-based positioning method by analyzing ranging performance characteristics and comparing positioning accuracy between the baseline closed-form (CF) approach and probabilistic grid-based positioning (PGP) across four anchor initialization configurations.

\subsection{Ranging Performance Analysis}
\label{subsec:ranging}

The ranging performance evaluation utilizes the complete dataset of 132,000 individual measurements from 12 UWB nodes deployed according to \cref{fig:arrange_conf}. Statistical analysis is categorized by propagation conditions (LOS/NLOS) to quantify measurement quality variations that directly impact collaborative positioning performance.

\begin{figure}[htbp]
    \centering
    \includegraphics[width=0.9\linewidth]{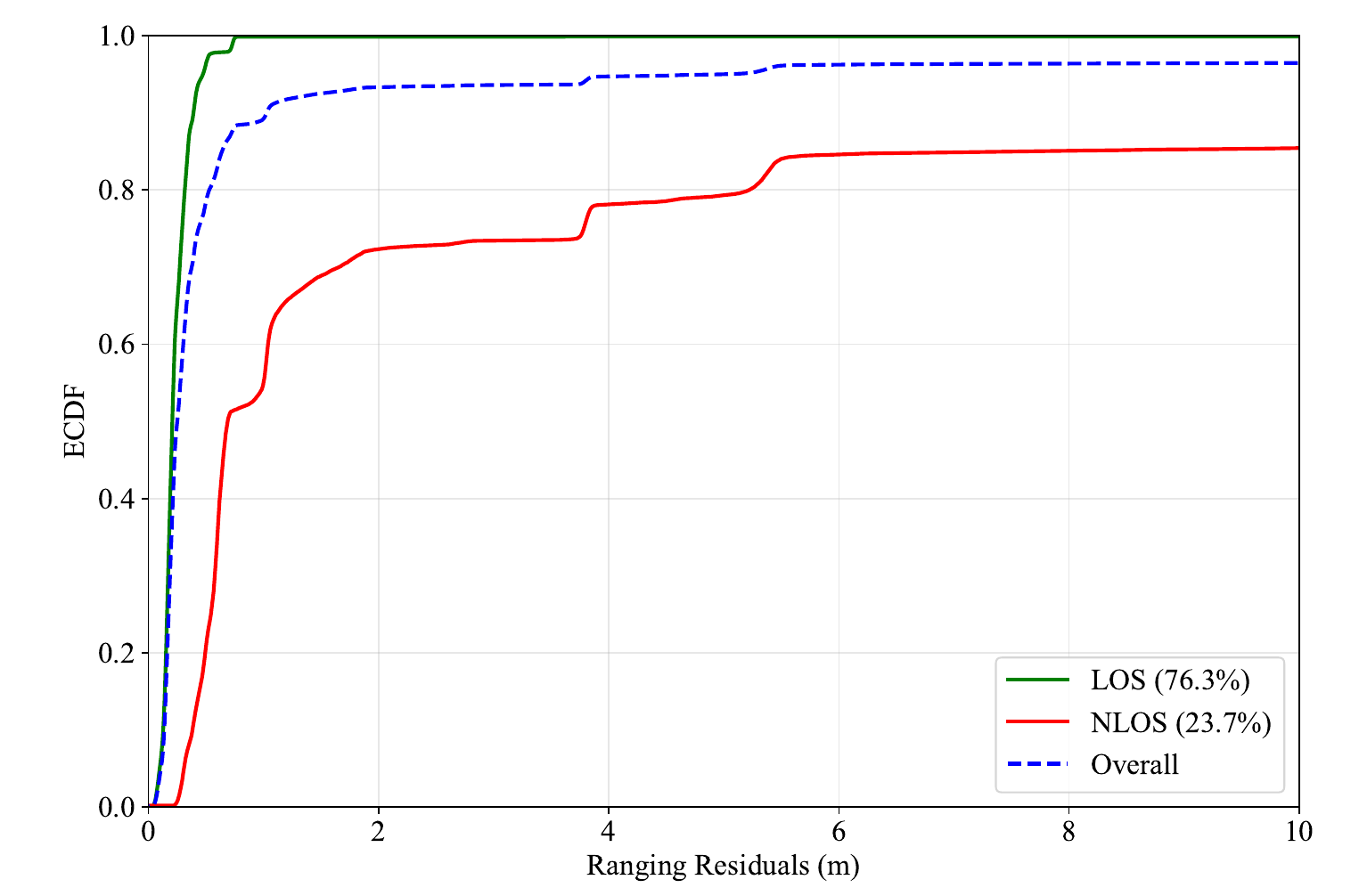}
    \caption{Ranging residual ECDF (LOS, NLOS, overall).}
    \label{fig:residual_ecdf}
    \vspace{-0.3cm}
\end{figure}

\begin{table}[htbp]
\caption{Ranging performance statistics.}
\label{tab:ranging_quality}
\centering
\begin{tabular}{lccc}
\toprule
\textbf{Metric} & \textbf{LOS} & \textbf{NLOS} & \textbf{Overall} \\
& $(\SI{76.3}{\percent})$ & $(\SI{23.7}{\percent})$ & $(\SI{100}{\percent})$ \\
\midrule
Mean $\mu$ (m) & 0.28 & 3.78 & 1.11 \\
Median (m) & 0.21 & 0.69 & 0.25 \\
Variance $\sigma^2$ (m) & 1.59 & 41.56 & 13.26 \\
\midrule
25th & 0.16 & 0.54 & 0.18 \\
50th & 0.21 & 0.69 & 0.25 \\
75th & 0.29 & 3.79 & 0.44 \\
\bottomrule
\end{tabular}
\end{table}

\Cref{tab:ranging_quality} presents ranging accuracy statistics across propagation conditions. LOS measurements achieve RMSE of $\SI{0.28}{\meter}$ with $\SI{0.21}{\meter}$ median error, exhibiting near-Gaussian characteristics. NLOS conditions show degraded performance with RMSE of $\SI{3.78}{\meter}$, reflecting multipath propagation and signal obstruction. Across both conditions, overall RMSE is $\SI{1.11}{\meter}$ with variance $\sigma^2 = \SI{13.26}{\meter\squared}$, indicating substantial measurement heterogeneity. The moderate NLOS error (3.78~m vs.\ $>$10~m for severe metal blockage~\cite{ridolfi_uwb_ML_2021}) represents realistic industrial deployment conditions with partial rather than complete signal occlusion.

\begin{figure}[htbp]
    \centering
    \includegraphics[width=1\linewidth, trim= 0cm .6cm 0cm 0cm, clip]{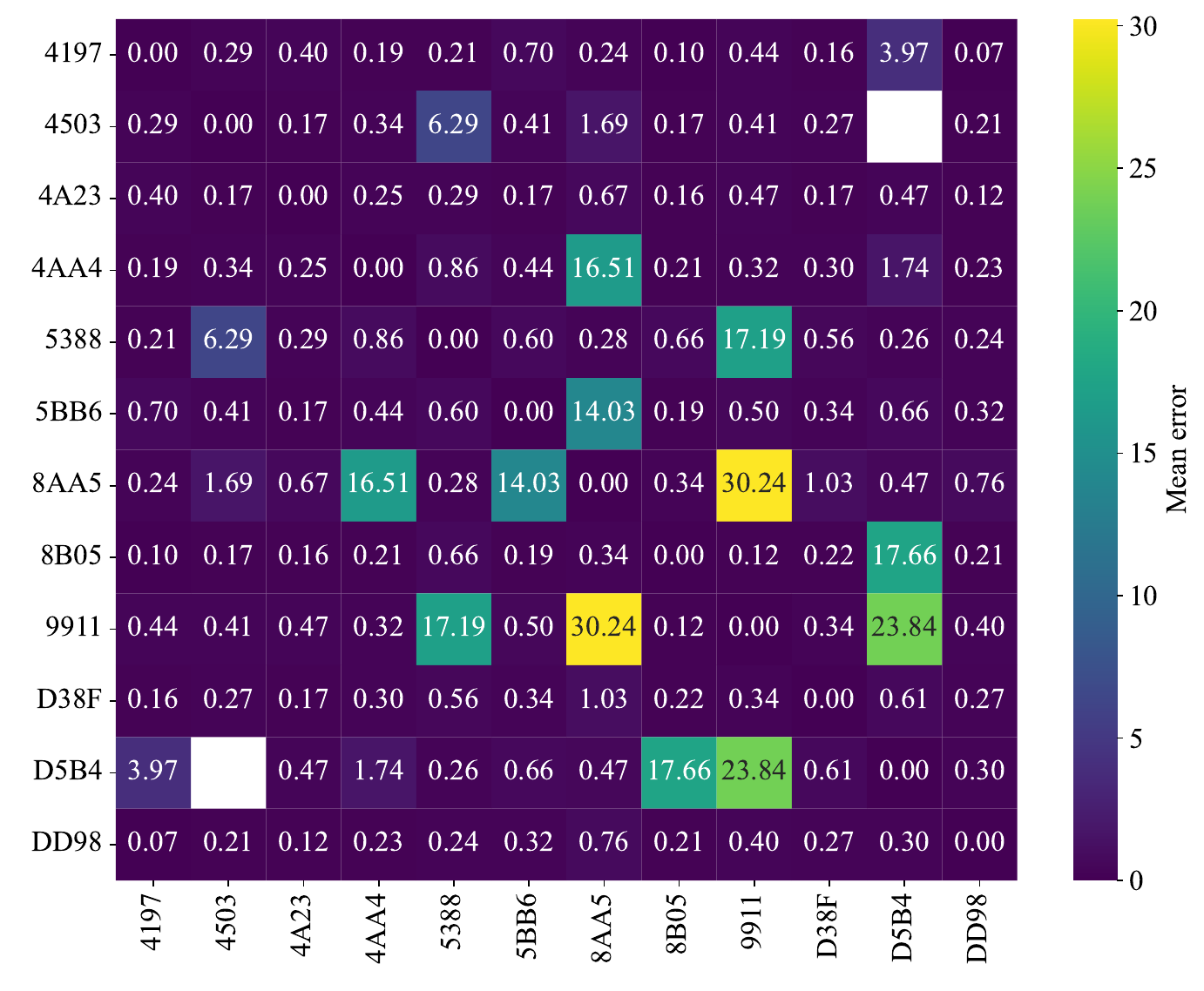}
    \caption{Ranging accuracy evaluation: Pairwise RMSE matrix.}
    \label{fig:rmse_matrix}
    \vspace{-0.3cm}
\end{figure}

The empirical cumulative distribution function (\cref{fig:residual_ecdf}) provides statistical characterization aligned with IPIN competition evaluation standards~\cite{Renaudin2019IPINFramework}, where the 75th percentile serves as the primary benchmark. LOS measurements exhibit tight error distribution (75th percentile: 0.29~m, 95th: 0.73~m), while NLOS exhibits heavy-tailed distribution with 75th percentile at 3.79~m. Still, 50\% of NLOS measurements remain below 0.69~m, preserving ranging utility under moderate occlusion. The overall 75th percentile (0.44~m) informed the grid-based algorithm's ranging noise parameter ($\sigma_r = 0.5$~m). The pairwise RMSE matrix (\cref{fig:rmse_matrix}) reveals systematic quality variations: nodes D5B4, 8AA5, and 9911 exhibit elevated error rates exceeding \SI{17}{\meter} RMSE, correlating with NLOS-dominated positions (\cref{fig:arrange_conf}).

\subsection{Measurement Availability Analysis}
\label{subsec:availability}

\begin{figure}[htbp]
\centerline{\includegraphics[width=1\columnwidth]{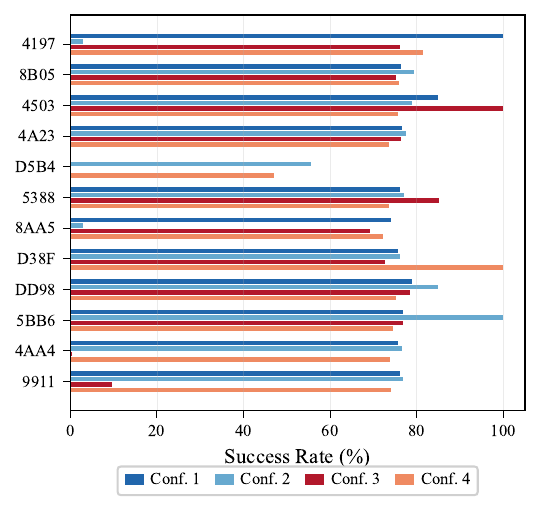}}
\caption{Measurement availability for the CF method across all the initialization configurations.}
\label{fig:availability}
\end{figure}

The availability analysis (\cref{fig:availability}) quantifies CF method sensitivity to anchor selection across all four configurations. Sensor D5B4 achieves zero positioning availability in both Conf.~1 and Conf.~3 due to missing ranges to initial anchors (\cref{fig:rmse_matrix}), yet recovers to $\SI{55.6}{\percent}$ and $\SI{47.0}{\percent}$ in Conf.~2 and Conf.~4, respectively. Similarly, sensors 4197 and 8AA5 drop from $\SI{100.0}{\percent}$ and $\SI{74.2}{\percent}$ in Conf.~1 to $\SI{3.0}{\percent}$ each in Conf.~2. The NLOS-contaminated configurations expose additional vulnerabilities: in Conf.~3, sensors 4AA4 and 9911 collapse to $\SI{0.4}{\percent}$ and $\SI{9.6}{\percent}$ respectively, despite maintaining $\approx\SI{75}{\percent}$ availability across all other configurations.

These configuration-specific failures are absent in the grid-based method, which maintains full positioning availability regardless of initialization choice.

These variations result from geometric constraints when insufficient connectivity occurs between the reference frame and target nodes. Cross-referencing with \cref{fig:rmse_matrix} reveals that positioning failures do not correspond to measurement quality degradation. Node D5B4 maintains ranging capabilities with multiple nodes despite elevated error rates with specific pairs. The NLOS configurations amplify this effect: the 4AA4/9911 failures in Conf.~3 arise because the initial anchor triangle (4503, 5388, 8AA5) has limited connectivity to these nodes, not because ranging quality deteriorates. The CF method's inability to exploit alternative measurement pathways reveals algorithmic rigidity. In contrast, the grid-based approach exhibits stronger tolerance to partial connectivity. The measurement update (\cref{eq:likelihood_uncertainty}) incorporates available nodes through $\mathcal{N}_i$, maintaining positioning capability even when individual measurements are missing.

The requirement for partial rather than complete distance matrices represents a fundamental methodological advantage for deployments with heterogeneous measurement availability.

\subsection{Positioning Performance Comparison}
\label{subsec:positioning}

\begin{figure}[htbp]
    \centering
    \includegraphics[width=\linewidth]{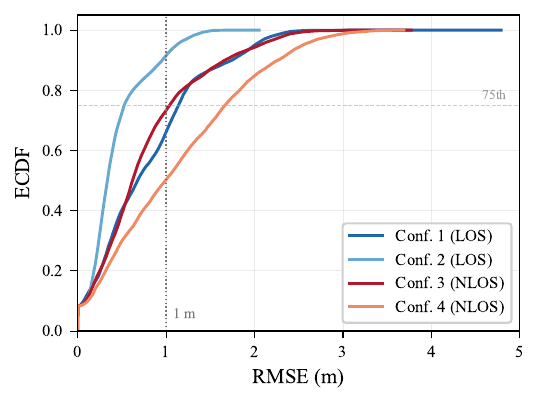}
    \caption{ECDF of positioning error for the PGP method across four initialization configurations. Vertical dotted line: 1\,m threshold. Horizontal dashed line: 75th percentile.}
    \label{fig:ecdf_pgp}
\end{figure}

\begin{figure}[htbp]
    \centering
    \includegraphics[width=\linewidth]{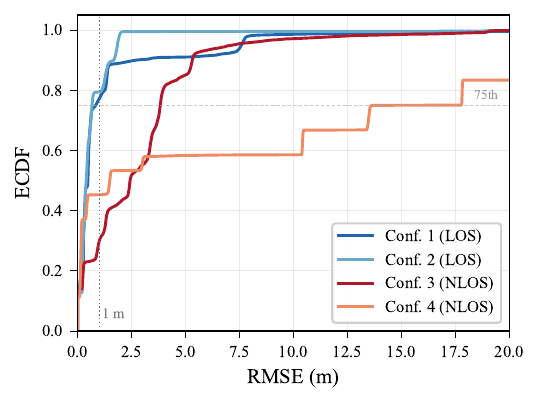}
    \caption{ECDF of positioning error for the CF method across four initialization configurations. Note the extended x-axis range (0--20\,m) compared to PGP (0--5\,m).}
    \label{fig:ecdf_cf}
\end{figure}

Positioning performance evaluation compares CF against PGP using identical measurement datasets across four initialization configurations (\cref{fig:arrange_conf}): two with pure LOS initialization (Conf.~1 and~2) and two with NLOS-contaminated initialization (Conf.~3 and~4), where 2 of 3 initial anchor links are NLOS-affected. This design isolates the effect of initialization propagation conditions on positioning accuracy. Summary statistics are provided in \cref{tab:positioning_summary}, with ECDFs shown in \cref{fig:ecdf_pgp,fig:ecdf_cf}.

\begin{table}[htbp]
\centering
\caption{Positioning error summary across four initialization configurations.}
\label{tab:positioning_summary}
\sisetup{round-mode=places, round-precision=2}
\begin{tabular}{@{}ll S[round-precision=2] S[round-precision=2] S[round-precision=2] S[round-precision=1,table-format=2.1] S[round-precision=2]@{}}
\toprule
Config & Method & {Median} & {p75} & {p95} & {${<}$1\,m} & {Mean} \\
 & & {(m)} & {(m)} & {(m)} & {(\%)} & {(m)} \\
\midrule
Conf.\ 1 (LOS) & PGP & 0.68 & 1.14 & 2.00 & 66.1 & 0.79 \\
Conf.\ 1 (LOS) & CF  & 0.47 & 0.83 & 7.59 & 77.2 & 1.28 \\
Conf.\ 2 (LOS) & PGP & 0.34 & 0.53 & 1.12 & 91.6 & 0.42 \\
Conf.\ 2 (LOS) & CF  & 0.40 & 0.66 & 1.84 & 79.4 & 0.67 \\
\midrule
Conf.\ 3 (NLOS) & PGP & 0.62 & 1.04 & 2.05 & 73.3 & 0.76 \\
Conf.\ 3 (NLOS) & CF  & 2.43 & 3.82 & 7.18 & 29.9 & 2.84 \\
Conf.\ 4 (NLOS) & PGP & 0.99 & 1.66 & 2.54 & 50.3 & 1.10 \\
Conf.\ 4 (NLOS) & CF  & 1.48 & 17.77 & 30.35 & 45.2 & 8.14 \\
\bottomrule
\end{tabular}
\vspace{-0.3cm}
\end{table}

\subsubsection{LOS Initialization Baseline}
Under pure LOS initialization, both methods achieve competitive performance. PGP attains median errors of 0.34--0.68\,m across the two LOS configurations, with 66--92\% of estimates below 1\,m (\cref{fig:ecdf_pgp}). The CF method achieves comparable median errors (0.40--0.47\,m) under these favorable conditions, though its 95th percentile for Conf.~1 (7.59\,m) already reveals vulnerability to outliers that the PGP method avoids (2.00\,m). The configuration-dependent variation in both methods reflects geometric dilution of precision from different anchor placements.

\subsubsection{NLOS-Contaminated Initialization}
The critical distinction emerges under NLOS-contaminated initialization (\cref{tab:positioning_summary}, lower half). The PGP method exhibits graceful degradation: median errors increase to 0.62--0.99\,m (from 0.34--0.68\,m under LOS), with the 75th percentile remaining below 1.66\,m. Notably, PGP under Conf.~3 (median 0.62\,m) performs comparably to PGP under Conf.~1 (median 0.68\,m), indicating that moderate NLOS contamination does not necessarily prevent accurate positioning.

In contrast, the CF method exhibits substantial performance collapse under NLOS initialization. Median errors increase to 1.48--2.43\,m, the 75th percentile reaches 17.77\,m for Conf.~4, and extreme outliers exceed 30\,m at the 95th percentile (\cref{fig:ecdf_cf}). The fraction of sub-meter estimates drops from 77--79\% (LOS) to 30--45\% (NLOS). This degradation reflects the deterministic nature of closed-form trilateration: NLOS bias in the initialization triangle propagates directly into the coordinate frame definition, systematically displacing all subsequent position estimates.

\subsubsection{Robustness Analysis}
The PGP--CF performance gap widens substantially under NLOS initialization. While both methods achieve similar median accuracy under LOS conditions, the mean error ratio (CF/PGP) increases from 1.6$\times$ under LOS to 3.7--7.4$\times$ under NLOS. This robustness stems from the grid-based method's probabilistic uncertainty propagation: NLOS-biased measurements create broader but representative probability distributions rather than catastrophic point failures. The measurement update (\cref{eq:likelihood_uncertainty}) weights grid cells by consistency across all available references, enabling the posterior to recover from initialization bias as additional measurements accumulate.

\section{Conclusion}
\label{sec:conclusion}

This paper presented grid-based uncertainty propagation for collaborative UWB self-calibration, 
addressing limitations of parametric methods that discard spatial distribution information 
when reducing anchor positions to point estimates. The proposed discrete Bayesian approach 
preserves probability mass throughout positioning, enabling reliable uncertainty quantification 
in NLOS-contaminated environments.

Experimental validation with 12 UWB nodes in a 44$\times$30\,m\textsuperscript{2} industrial environment generated 132,000 ranging measurements. LOS conditions achieved 0.28\,m mean ranging error versus 3.78\,m for NLOS. The grid-based approach achieved consistent sub-meter positioning accuracy under LOS initialization, while baseline closed-form methods exhibited substantial sensitivity to initialization selection. 

Under NLOS-contaminated initialization, the grid-based approach maintained sub-meter median accuracy (0.62--0.99\,m) while the closed-form baseline exhibited order-of-magnitude degradation with median errors up to 2.43\,m and extreme outliers exceeding 30\,m, confirming the practical advantage of probabilistic uncertainty propagation for deployments where LOS initialization cannot be guaranteed.

Future research directions include quantitative comparison with MDS and robust localization
baselines, as well as distributed processing architectures for large-scale deployments. Multi-technology integration combining WiFi, BLE, and 5G within the probabilistic framework could enhance coverage. Machine learning may further enable automated measurement quality assessment.

\section*{Data Availability Statement}
The presented dataset, comprising 132,000 rangings with ground truth, LOS/NLOS classifications, and node configurations, will be made publicly available through the institutional repository of TU Dresden University of Technology following publication, facilitating reproducibility and future research.



\section*{Acknowledgment}
The authors acknowledge Deveritec GmbH for providing updated firmware (under proprietary licensing) enabling fully meshed measurements and MRK Management Consultants GmbH for facility access. Language editing of the revised manuscript was assisted by a generative AI tool (Claude, Anthropic). All scientific content, analysis, and interpretation remain solely the work of the authors.

\printbibliography

\end{document}